\documentclass[prl,twocolumn,showpacs,preprintnumbers,amsmath,amssymb]{revtex4}

\usepackage{epsfig}
\usepackage{graphicx}
\usepackage{dcolumn}
\usepackage{bm}

\newcommand{\vf}{v_{\rm F}}

\newcommand{\ef}{\epsilon_{\rm F}}

\newcommand{\kf}{k_{\rm F}}

\begin{document}

\title{Intervalley scattering, long-range disorder,
 and effective time reversal symmetry breaking in
  graphene.}
\author{A. F. Morpurgo}
\affiliation{Kavli Institute of Nanoscience, Delft University of Technology,
  Lorentzweg 1, 2628 CJ Delft, The Netherlands}
\author{F. Guinea}
\affiliation{Instituto de Ciencia de Materiales de Madrid. CSIC. Cantoblanco. E-28015
Madrid. Spain.}

\begin{abstract}
We discuss the effect of certain types of static disorder, like that induced
by curvature or topological defects,  on the quantum
correction to the conductivity in graphene. We find that when the intervalley
scattering time is long or comparable to $\tau_{\phi}$, these defects can
induce an effective time reversal symmetry breaking of the hamiltonian
associated to each one of the two valleys in graphene. The phenomenon suppresses the magnitude of the quantum correction to the conductivity and may result in the complete absence of a low field magnetoresistance, as recently found
experimentally.  Our work shows that a quantitative description of weak localization in graphene must include the analysis of new regimes, not present in conventional two dimensional electron gases.
\end{abstract}

\maketitle
{\em Introduction.} Graphene provides a two-dimensional electron
system that is distinctly different from the two-dimensional
electron gases (2DEGs) hosted in common semiconducting
heterostrutures. The low energy electronic states of graphene can be
described by two sets of two-dimensional spinors associated to two
independent points at the corner of the Brillouin zone ($K$ and
$K'$). In the absence of short range potentials the states
associated to the different points are not coupled and, in the long
wavelength limit, they satisfy the two dimensional Dirac equation.
Theoretical calculations based on these assumptions\cite{PGN06}
reproduce the unconventional quantization of the quantum Hall
conductance observed experimentally \cite{Netal05,Zetal05b} (see
also\cite{Ketal03}). It is expected that, beside the quantum Hall
effect, other phenomena should manifest unusual characteristics as
compared to more conventional 2DEGs.

It appears that one of these phenomena is the weak-localization
correction to the conductivity\cite{AKLL80,LK82,B84,K84}. In  the
presence of time-reversal symmetry, the suppression of
weak-localization at low magnetic fields produces a negative
magnetoresistance, ubiquitous in metallic conductors. The magnitude
of the weak-localization correction is of the order of $e^2/h$ and
it is determined by two characteristic time scales: the phase
coherence time $\tau_{\phi}$ and the elastic scattering time $\tau$.
In the graphene samples that exhibit very clear QHE, however, no
low-field magnetoresistance is observed. In these samples,
nevertheless, the observation of high-order quantum Hall plateau
indicates that phase coherent propagation of electrons occurs at
least on a distance of several hundreds nanometers, which
corresponds to the estimated elastic mean-free path, so that a
magnetoresistance originating from the suppression of
weak-localization is expected.

It has been realized by Suzuura and Ando\cite{SA02} that the quantum
correction to the conductivity in graphene can differ from what is
observed in conventional 2DEGs (see also\cite{K05}). This is due to
the pseudo-spin associated to the solutions of the Dirac equation,
that in conjunction with the nature of elastic scattering in
graphene may change the sign of the localization correction, and
turn weak-localization into weak-antilocalization (even in the
absence of spin-orbit interaction). The phenomenon crucially depends
on the inter-valley scattering time $\tau_{\rm iv}$, i.e. the
characteristic times that it takes for a charge carrier to be
scattered from one to the other $K$-point in graphene. Specifically,
if $\tau_{\rm iv} \gg \tau_{\phi}$ weak antilocalization is expected,
resulting in a positive magnetoresistance at low field; if
$\tau_{\rm iv} \ll \tau_{\phi}$, conventional weak localization should
be observed. Even though these conclusions are at odds with
experimental findings, they are important as they clearly illustrate
the relevance of inter-valley scattering.

Here, we show that not only weak localization and antilocalization
can appear in graphene depending on the nature of elastic
scattering, but also that the quantum correction to the conductivity
can be entirely suppressed due to time-independent potential slowly
varying in space. These static potentials can result in the
effective breaking of time reversal symmetry of electronic states
around each $K$-point when $\tau_{iv} > \tau_{\phi}$. They originate
from defects that are realistically present in graphene, such as
long-range distortions induced by topological lattice defects
(disclinations and dislocations), non-planarity of the graphene
layers, and slowly varying random electrostatic potentials that
break the symmetry between the two sublattices of graphene. We
estimate the effect of each one of these defects in terms of a
characteristic time, which acts as a cut off for the time-reversed
trajectories of electrons responsible for weak-localization
phenomena. If the characteristic time associated to one of these
mechanisms is much shorter than $\tau_{\phi}$, weak-localization is
largely suppressed. This explains the absence of weak-localization
in the first magneto-transport experiments in graphene.

We emphasize that the potentials that we consider, being static, do
not actually break time reversal symmetry in graphene. However, in
the presence of these potentials, time reversal symmetry connects
electronic states associated to the two different $K$-points. If the
electron dynamics is such that electrons cannot be transferred from
one to the other $K$-point within their phase coherence time (i.e.,
if $\tau_{iv} > \tau_{\phi}$), the contribution to interference due
to time-reversed trajectories (responsible for weak-localization) is
suppressed\cite{note1}.

{\em The model.}
We use the continuum approximation to the band structure near the $K$ and
$K'$ points of the Brillouin Zone of graphene. The wavefunctions can be
expressed as a two component spinor, $[ \Psi_{A , K , \uparrow} ( {\bf
  \vec{r}} ) , \Psi_{B , K , \uparrow} ( {\bf  \vec{r}} ) ]$, where $A$ and
$B$ stand for the two sublattices of the honeycomb structure. Other spinors
can be defined for the $K'$ point and the down spin orientation. The
curvature of the graphene sheet can be included by generalizing the ordinary
derivative operator to the covariant derivative\cite{GGV92,GGV93b}, following
the standard procedure used for spinor fields\cite{BD82}. These effects are
described below by the spin connection
operator $\hat{\bf \Sigma}_{\rm curv}$ which breaks the effective time
reversal symmetry around each $K$ point. Similarly, certain topological
lattice defects imply the existence of rings with an odd number of Carbon
atoms. A dislocation has a pentagon at its core, and a dislocation requires a
pentagon-heptagon pair. An odd numbered ring of Carbon atoms implies that the
two sublattices are interchanged when traversing a path which encloses it. This needs to be
taken into account in the continuum description, irrespective of the
smoothness of the induced distortion. It can be described by a non Abelian
gauge potential $\hat{\bf \Sigma}_{\rm def}$ \cite{GGV92,GGV93b}.
It has been shown that its inclusion allows for an accurate description
of the electronic spectrum in curved graphene sheets, such as
fullerenes\cite{GGV92,GGV93b}, or graphitic cones\cite{LC00b}.
Note that this gauge field interchanges the two sublattices and
the $K$ and $K'$ points. This fact, however, does not prevent the
description of the system in terms of two sets of spinorial wavefunctions
which obbey equivalent Dirac equations. In fact, using the transformation:
\begin{eqnarray}
\tilde{\Psi} _{A {\cal K} {\bf \vec{k}} s} ( {\bf \vec{r}} ) &=
&\Psi_{A K {\bf \vec{k}} s} ( {\bf \vec{r}} ) +
i \Psi_{B K' {\bf \vec{k}} s} ( {\bf \vec{r}} ) \nonumber \\
\tilde{\Psi}_{B {\cal K'} {\bf \vec{k}} s} ( {\bf \vec{r}} ) &= &\Psi_{B K' {\bf
  \vec{k}} s} ( {\bf \vec{r}} ) - i \Psi_{A K {\bf \vec{k}} s} ( {\bf
\vec{r}} )
\label{transformation}
\end{eqnarray}
the model can be reduced to two independent hamiltonians:
\begin{widetext}
\begin{eqnarray}
  {\cal H}_{{\cal K} , s} &\equiv &\left( \begin{array}{cc} \Delta ( {\bf
  \vec{r}} ) &\vf ( i \partial_x + \partial_y ) +
  {\hat{\bf A}}_x + i {\hat{\bf A}}_y + \hat{\bf \Sigma}_{\rm curv} +
  \hat{\bf \Sigma}_{\rm def} \\ \vf ( i \partial_x -
  \partial_y ) + {\hat{\bf A}}_x -  i {\hat{\bf A}}_y + \hat{\bf \Sigma}_{\rm
  curv} + \hat{\bf \Sigma}_{\rm def}&- \Delta ( {\bf \vec{r}} )
  \end{array} \right) \nonumber \\
  {\cal H}_{{\cal K'} , s} &\equiv &\left( \begin{array}{cc} \Delta ( {\bf
  \vec{r}} ) &\vf ( i \partial_x - \partial_y ) +
  {\hat{\bf A}}_x - i {\hat{\bf A}}_y + \hat{\bf \Sigma}_{\rm curv}
- \hat{\bf \Sigma}_{\rm def} \\ \vf ( i \partial_x +
  \partial_y ) + {\hat{\bf A}}_x +  i {\hat{\bf A}}_y +
\hat{\bf \Sigma}_{\rm curv}
- \hat{\bf \Sigma}_{\rm def} &- \Delta ( {\bf
   \vec{r}} ) \end{array} \right)
\label{hamil}
\end{eqnarray}
\end{widetext}
where $\hat{\bf A}$ is the ordinary vector potential. $\Delta ( {\bf
  \vec{r}} ) = V_{\rm A} ( {\bf \vec{r}} ) - V_{\rm B} ( {\bf \vec{r}} )$
measures the difference of the electrostatic potential in the two
sublattices and originates from the environment around the graphene
layer (e.g., charges located at random position in the substrate
supporting graphene).

A description of quantum interference in terms states associated to
two decoupled K-points separately is correct if $\tau_{\rm iv} \gg
\tau_{\phi}$. In this regime, the defects described by three terms
$\hat{\bf \Sigma}_{\rm def} , \hat{\bf \Sigma}_{\rm curv}$ and
$\Delta ( {\bf \vec{r}} )$ break the effective time reversal
symmetry around the ${\it K}$ and ${\it K'}$ points, which can be
written as the complex conjugation operator times
$\sigma_y$\cite{MF04}. We will treat separately the effect of the
different defects by analyzing the dynamical phase that
quasi-classical trajectories in real space acquire in their
presence. We will confine ourselves to the qualitative analysis of
this problem, but we note that a formal description of weak
-(anti)localization in terms of quasi-classical trajectories
\cite{LK82,K84,B84}can be made equivalent to diagrammatic
approaches\cite{CS86}.

\begin{figure}
\begin{center}
\includegraphics*[width=5cm,angle=-0]{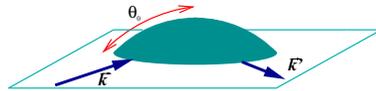}
\caption{Geometry of the scattering process analyzed in the text. An impurity
potential is placed at the center of a spherical bump in a flat sheet.}
\label{bump}
\end{center}
\end{figure}
{\em Curvature effects.}
Scattering is modified in a curved surface. We analyze the geometry
sketched in Fig.[\ref{bump}]. An incoming plane wave moves into a spherical
cap parametrized by the angle $\theta$.  The wave is scattered by a defect
within the bump. We place the defect at the center of the region, in order to
simplify the calculation. In flat space, the incoming and outgoing waves can
be written as:
\begin{equation}
\begin{array}{lr} \Psi_{\bf \vec{k}} \equiv \left( \begin{array}{c} 1 \\ e^{i
      \phi_{\bf \vec{k}}} \end{array} \right) e^{i {\bf \vec{k}} {\bf
      \vec{r}}}
& \Psi_{\bf \vec{k}'} 
\equiv \left( \begin{array}{c} 1 \\ e^{i
      \phi_{\bf \vec{k}'}} \end{array} \right)  e^{i {\bf \vec{k}'} {\bf
      \vec{r}}}
\end{array}
\label{plane_waves}
\end{equation}
where $\phi_{\bf \vec{k}}$ determines the direction of ${\bf \vec{k}}$.
The scattering amplitude due to a local potential, $\hat{V} \equiv 
V \delta ( {\bf  \vec{r}} )$  is $A = \langle {\bf \vec{k}} | \hat{V} | {\bf
\vec{k}'} \rangle \propto V [ 1 + \cos ( \phi_{\bf \vec{k}} - \phi_{\bf \vec{k}'}
) ]$.

\begin{figure}
\begin{center}
\includegraphics*[width=5cm,angle=-90]{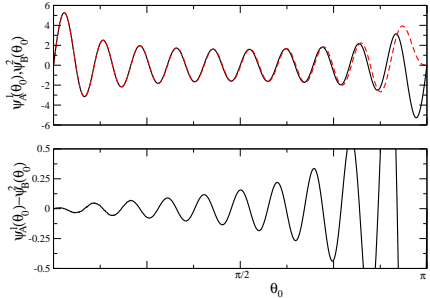}
\end{center}
\caption{Wavefunctions within the spherical cap shown in
  Fig.[\protect{\ref{bump}}] for ${\cal l}=20$. These two solutions, in flat
  space, are related by the effective time reversal symmetry defined in each valley.}
\label{amplitude_bump}
\end{figure}

We calculate the scattering amplitude in the geometry shown in
Fig.[\ref{bump}] by matching the plane waves in eq.(\ref{plane_waves}) to
solutions inside the spherical cap. The wavefunctions can be
computed analytically for energies $\epsilon_l = \vf / R \sqrt{{\bf l}
  ( {\bf l}+1 )- 2}$\cite{GGV92}. The quasiclassical limit corresponds to
${\bf l} \gg
1$. The main difference between the solutions inside the cap and those in a
plane, eq.(\ref{plane_waves}), is that the values of the modulii of the two components of the
Dirac spinors are not equal. Because of the radial symmetry of the
problem, we only need the solutions inside the cap with $m=0$. The relevant
solutions, for a given valley, can be written as:
\begin{equation}
\begin{array}{lr} \Psi_{1} \equiv \left( \begin{array}{c} \psi_A^1 ( \theta )
      \\ \psi_B^1 ( \theta ) e^{- i \phi} \end{array} \right)
& \Psi_{2} 
\equiv \left( \begin{array}{c}  \psi_A^2 ( \theta ) e^{i \phi} \\ \psi_B^2 (
    \theta ) \end{array} \right) 
\end{array}
\end{equation}
Matching these solutions to those in eq.(\ref{plane_waves}) outside the cap,
the scattering amplitude can be written as:
\begin{equation}
A  \propto V [ | {\psi_A^1} ( \theta_0 )^2 + {\psi_B^2} ( \theta_0 )^2 + 2
{\psi_A^1} ( \theta_0 ) {\psi_B^2} ( \theta_0 )
\cos ( \phi_{\bf \vec{k}} - \phi_{\bf \vec{k}'}) ]
\end{equation}
where $\theta_0$ is the angle which defines the shape of the cap. The
deviation from scattering in a plane is determined by the difference
${\psi_A^1} ( \theta_0 ) - {\psi_B^2} ( \theta_0 )$, which is maximum when
${\psi_A^1} ( \theta_0 ) , {\psi_B^2} ( \theta_0 ) \approx 0$, as shown in
Fig.[\ref{amplitude_bump}].
 
The first zeros of the functions ${\psi_A^1} ( \theta_0 ) , {\psi_B^2} ( \theta_0 )
\approx 0$ lie at $\theta_0 \sim {\bf l}^{-1}$. Hence, we expect deviations from
antilocalization for Fermi energies such that $\theta_0 \sim \ef R / \vf$,
that is, $\kf R \sim 1$. In a sample with height fluctuations of order $h$
oner a length $l$, we have $R^{-1} \sim h^2 / l$, so that antilocalization
effects will be suppressed for $( \kf h^2 ) / l \sim 1$.

{\em Gauge field induced by lattice defects.}
As mentioned above, a local rotation of the axes leads to the existence of a
gauge field, $\hat{\bf \Sigma}_{\rm def}$. In terms of the lattice strain,
$u_x ( {\bf r} ) , u_y ( {\bf r} )$, this rotation is given by\cite{GGV01}:
\begin{equation}
\theta ( {\bf r} ) \approx \frac{\partial_y u_x - \partial_x u_y}{2}
\label{theta}
\end{equation}
and $\hat{\bf \Sigma}_{\rm def}$ in eq.(\ref{hamil}) becomes:
\begin{equation}
\hat{\bf \Sigma}_{\rm def} \equiv \frac{1}{2} \nabla \theta ( {\bf r} )
\label{gauge}
\end{equation}
We consider first the effect of gauge field induced by a finite distribution of
dislocations, with random Burgess vectors ${\bf b}_i$, and average distance
$d$, on a quasiparticle moving around a closed loop of length $l$.

At sufficiently long distances from an edge dislocation with Burgess vector
${\bf \vec{b}}$, eq.(\ref{theta}) gives
$\theta ( {\bf r} ) \propto {\bf  \vec{b}}{\bf \vec{r}} / | {\bf r} |^2 $.
The local rotations average to zero when the loop encloses
completely the dislocations. A finite contribution is obtained from the
dislocations whose core is near the trajectory of the particle.
The quasi-classical width of the trajectory of the particle is of order
$\kf^{-1}$, where $\kf$ is the Fermi wavevector. Hence, the number of
dislocations which contribute to change the phase around the loop is of order
$( \kf^{-1} d ) / l^2$, where $l$ is the length of the loop and $d$
is the average distance between dislocations.

To estimate the effect of the rotation on the phase of the trajectories
traversing the loop, we assume that the Burgess vector of the dislocations
is distributed randomly. Using the central limit theorem, we find that the
rotation associated to the loop becomes non negligible when
$[ ( \kf^{-1} d ) / l^2 ]^2 \gg 1$. This rotation leads to a phase of order
$ \pi$, which dephases randomly trajectories traversed clock- and
anticlockwise,
which suppresses the quantum interference correction to the conductivity.
From the above inequality, the time scale at which these effects are relevant is:
\begin{equation}
\tau_{\rm gauge}^{-1} \sim \frac{\vf}{\sqrt{\kf^{-1} d}}
\end{equation}

{\em Potential gradients.} The potential $\Delta ( {\bf \vec{r}} )$
in eq.(\ref{hamil}) represents the difference in potentials at the
two sites of the unit cell. Physically, this potential originates
from charges located in the substrate supporting the graphene
layer\cite{note2}. An asymmetry between the two sublattices
 arises from slowly varying potentials,
that can be written as the sum of a smooth term which is the same
within each unit cell, and a small contribution which breaks the
equivalence of the two sites in the unit cell.  The second part can
be written as $\Delta ( {\bf \vec{r}} ) \approx {\bf \vec{c}} \nabla
V ( {\bf \vec{r}} )$,
 We assume this potential to be slowly varying, so that:
\begin{equation}
V ( {\bf \vec{q}} ) \approx \left\{ \begin{array}{lr}
V_0 & | {\bf \vec{q}} | \ll q_c \\  0 &| {\bf \vec{q}} | \gg q_c \end{array}
\right.
\label{disorder}
\end{equation}
where $q_c$ is a momentum cutoff such that $q_c \ll a^{-1}$, where $a$ is the
lattice constant.  We are interested in momentum transfers $| {\bf
  \vec{q}} | \sim \kf$, where $\kf \ll a^{-1}$ for typical
dopings. Eq.(\ref{disorder}) implies that:
\begin{equation}
\Delta ( {\bf \vec{q}} ) \approx \left\{ \begin{array}{lr} {\bf \vec{q}} {\bf
      \vec{c}} V_0 &| {\bf \vec{q}} | \ll q_c \\ 0 &| {\bf \vec{q}} | \gg q_c
      \end{array} \right.
\end{equation}
The elastic scattering time can be obtained from $V_{\rm disorder}$ using
Fermi's golden rule:
\begin{equation}
\tau_{\rm elastic}^{-1} \approx \frac{V_0^2 \nu ( \ef )}{a^2}
\end{equation}
where $\nu ( \ef ) = | \ef | / ( \pi \vf^2 )$ is the density of states at the
 Fermi energy.

The potential $\Delta ( {\bf \vec{r}} )$
breaks the effective
time reversal symmetry around each $K$ point. We can estimate the
inverse time at which this operator induces a significant change in the
wavefuction of the spinor applying again Fermi's golden rule to the symmetry breaking potential:
\begin{equation}
\tau_{\rm grad}^{-1} \approx \frac{V_0^2 \nu ( \ef )}{a^2} | {\bf \vec{c}} |^2 \kf^2
\sim \tau_{\rm elastic}^{-1} n_{\rm elec}
\end{equation}
where $n_{\rm elec} \sim ( \kf a )^2$ is the number of carriers per unit cell (note that the effect can be larger close to the sample edges where the disorder is also larger). In typical experiments, $n_{\rm el} \sim 10^{-4} - 10^{-2}$. Thus, there exists a range of energies or temperatures, $\tau_{\rm grad}^{-1} \lesssim T \lesssim \tau_{\rm el}^{-1}, \tau^{-1}_\phi \ll T$, where the effective time reversal
symmetry around each ${\it K}$ is not broken by $\Delta ( {\bf \vec{r}} )$.

{\em Conclusions.} Our analysis illustrates that the behavior of the
quantum correction to the conductivity in graphene is much richer
than what was anticipated in the work of Suzuura and Ando
\cite{SA02}. We conclude that the inter-valley scattering time
$\tau_{iv}$ and the phase coherence time $\tau_{\phi}$ are not the
only important time scales. An additional time scale describing the
effective time reversal symmetry breaking is present, which can
cause a complete suppression of weak (anti)localization. This time
scale depends on specific defects present in the graphene samples,
which leads to the prediction that large differences in the quantum
correction to the conductivity measured on different samples should
be expected. This is striking, since in all metallic conductors
studied in the past weak (antilocalization manifests as a robust and
very reproducible phenomenon.

Furthermore, the physical understanding provided by our analysis,
which does not rely on any detailed assumption, enables us to draw
additional conclusions. For instance, we expect that it will be
easier to observe weak-localization in narrow graphene samples,
since scattering at the edges couple states at the two different
K-points. We also expect that an effective time reversal symmetry
breaking similar to the one discussed here should occur more in
general, in systems with two or more degenerate valleys with
topological defects similar to those considered here. This is the
case, for instance, for a graphene bilayer\cite{MF06,Ketal06}. Note,
however, that contrary to individual graphene layers, in graphene
bilayers the relative phases of the two components in the spinor of
the momentum eigenfunctions are twice those in graphene. Therefore,
in sufficiently clean samples (where defects are not sufficient to
induce an effective time reversal symmetry breaking) the usual
negative magnetoresistance is expected, and there should be no weak
antilocalization \`a la Suzuura-Ando.

We gratefully acknowledge very useful discussions with A. Geim, K. Novoselov,
P. Kim, B. Trauzettel, A. Castro Neto, N. M. R. Peres, M. Katsnelson,
L. Vandersypen, H. Heersche, and P. Jarillo-Herrero. F. G. acknowledges
funding from MEC (Spain) through grant FIS2005-05478-C02-01 .

{\em Note added.} Since this work was posted, three related works have
appeared\cite{Metal06,Metal06b,NM06} dealing with the same
topic. Ref.\cite{Metal06} presents experimental results consistent with our
work, as well as a theoretical explanation along similar lines. The analysis of weak
localization effects in\cite{NM06} is also consistent with our results. An
alternative explanation is discused in\cite{Metal06b}. 
\bibliography{graphite0_1}

\begin{thebibliography}{24}
\bibitem[{not({\natexlab{b}})}]{note2}
\bibinfo{note}{On average, the potential generated by charges in the substrates
  is identical for both the A and B sublattices. However, for any particular
  graphene sample the potential will deviate from the average, and will not be
  identical on the two sublattices.}
\end{thebibliography}
\end{document}